# Modulated optical reflectance measurements on La$_{2/3}$Sr$_{1/3}$MnO$_3$ thin films

Laurence Méchin, Stéphane Flament

GREYC (CNRS - UMR 6072), ENSICAEN and University of Caen, 6 boulevard du Maréchal Juin, 14050 Caen cedex, France

Andy Perry, Darryl P. Almond

Materials Research Centre, University of Bath, Bath BA2 7AY, UK

Radoslav A. Chakalov

School of Physics and Astronomy, University of Birmingham, Birmingham B15 2TT, UK

Corresponding author :

Dr Laurence Mechin

GREYC (UMR 6072), ENSICAEN, 6 bd Maréchal Juin, 14050 CAEN cedex, France

Tel: +33 2 31 45 26 92

Fax: +33 2 31 45 26 98

E-mail : lmechin@greyc.ensicaen.fr

**ABSTRACT**

The modulated optical reflectance (MOR) measurement technique was applied to colossal magnetoresistive materials, in particular, La$_{2/3}$Sr$_{1/3}$MnO$_3$ (LSMO) thin films. The contactless measurement scheme is prospective for many applications spanning from materials characterization to new devices like reading heads for magnetically recorded media. A contrasted room temperature surface scan of a 100 μm wide 400 μm long bridge patterned into LSMO film provided preliminary information about the film homogeneity. Then the temperature was varied between 240 and 400 K, i.e. through the ferromagnetic to paramagnetic transition. A clear relation between the MOR signal measured as function of the temperature and the relative derivative of the resistivity up to the Curie temperature was observed. This relationship is fundamental for the MOR technique and its mechanism was explored in the particular case of LSMO. Analysis in the framework of the Drude model showed that, within certain conditions, the measured MOR signal changes are correlated to changes in the charge carrier concentration.

# 1. INTRODUCTION

The Modulated Optical Reflectance (MOR) technique is based on the measurement of the optical reflectance variations of a sample subjected to periodic heat stimuli [1, 2]. Since optical and electrical properties are closely interrelated, the method has been used for qualitative characterization of the electronic transport properties of crystalline and ion-implanted semiconductors [3, 4], bulk metals and thin metallic layers [5] or for the control of the quality of wafer surface preparation [6]. MOR is a rapid, contactless and non destructive technique, and can offer high spatial resolution (few µm and better). It is now widely accepted in semiconductor manufacturing process as a standard method for real time monitoring of ion implantation [7]. The potential of MOR for device quality assessment of high-temperature superconducting (HTS) thin-film devices was also demonstrated [8, 9]. It was used in particular for room-temperature non destructive evaluation of the surface homogeneity of YBCO thin films for microwave applications [8, 10].

Perovskite manganese oxides of general formula $RE_{1-x}M_xMnO_3$ (RE = rare earth, M = Ca, Sr, Ba, Pb) have raised renewed interest because they exhibit a rich variety of electronic and magnetic properties, including colossal magnetoresistance (CMR), charge ordering and phase segregation [11, 12, 13]. They can be prepared in thin film form using the deposition techniques previously developed for HTS films [14, 15]. Because of the large magnetoresistance effect, the strong spin polarization at the Fermi level but also the metal-to-insulator transition, these oxides may find important applications in magnetoresistive devices such as magnetic random access memory and sensors [16]. Among the variety of manganite compositions, $La_{2/3}Sr_{1/3}MnO_3$ (LSMO) thin films are good candidates for device applications because of their high Curie temperature (367 K for bulk LSMO).

The purpose of the experimental work presented in this paper was to study the applicability of the MOR technique to LSMO thin films and to determine its potential both for materials characterization and for new device development. In the former case, the advantage of the contactless measurement scheme could be expanded even further – a good spatial separation of the optical system from the sample could prove being invaluable for situations where the sample must be subjected to special conditions e.g. high magnetic fields. In the other case, the MOR technique could be used as reading tool of magnetic recording media which is not based on the Kerr effect. For both of these purposes, a better knowledge of the mechanism behind the MOR signal generation and its relation to the LSMO film resistance is necessary.

The sample preparation and the experimental conditions are described in section 2. We explore the basic application of the technique – for materials homogeneity assessment. A map of the MOR response all over the surface of a 100 μm wide and 400 μm long bridge is presented in section 3. To further investigate the origin of the MOR signal we vary the temperature between 240 and 400 K (through the ferromagnetic to paramagnetic transition). We find that the MOR signal is proportional to the resistivity temperature coefficient. The manganites belong to the group of highly correlated electron systems where charge, spin, and lattice degrees of freedom are intimately related. Although models for the conductivity in such systems are not yet well established, we present an analysis of the results and give a tentative explanation about the origin of the measured MOR signals using simple considerations in section 4.

## 2. SAMPLE PREPARATION AND EXPERIMENTAL CONDITIONS

A 200 nm thick LSMO thin film was deposited by pulsed laser deposition from a stoichiometric target onto a (100) $SrTiO_3$ single crystal substrate. To avoid deviations of the ablated material chemical composition due to the non-uniform laser beam intensity profile, an external optical beam intensity homogeniser was used. The laser energy density was 1.5 J/cm², the target-to-substrate distance - 65 mm, the oxygen pressure - 0.700 Torr and the substrate temperature - 780 °C. These parameter values were found optimal for producing single-crystalline films (as judged by the X-Ray Diffraction Patterns) with smooth surface. The X-Ray Diffraction study indicated a high (100) orientation of the LSMO film. The roughness was measured by Atomic Force Microscopy and the RMS value was 2.21 nm for a 14 μm by 14 μm image. A 500 nm thick gold layer was deposited by RF sputtering to prepare contact pads for low resistive four-probe connections. The LSMO thin film was patterned by UV photolitography and argon ion etching into 400 μm long × 100 μm wide bridges. The corresponding contact pads in the patterned film and the sample holder were finally connected using aluminium wires by ultrasonic bonding.

Figure 1 is a scheme of the MOR system used in this study, which is similar to that described by Opsal et al. [2] and previously used in Ref. 8 and 9. An Adlas DPY 315C frequency doubled semiconductor laser emitting a 533 nm beam was used as the pump laser and a Uniphase 1350P He/Ne laser provided the 632.8 nm probe beam. Both laser beams were focused onto the sample using a conventional 10× microscope objective lens. The resulting focal spot was estimated to have a diameter of about 15 μm. The pump and the probe laser beam powers coming onto the sample surface were measured to be 12.6 mW (without any filter) and 2.80 mW, respectively.

The pump laser was square-wave modulated by an acousto-optic cell at $f_c = 1$ kHz. Its purpose is to locally modulate the temperature of the sample film. The probe beam was directed to the pump laser heated spot. After reflection, it was separated and treated independently of the reflected pump beam using a polarizing beam splitter, a quarter waveplate and a He/Ne wavelength selective filter. The probe beam modulation was detected by a Hamamatsu 133BQ photodiode and monitored using an EG&G 5210 lock-in amplifier, thus providing an electrical signal proportional to the reflectivity modulation $\Delta R$ at each point of the sample. The whole system was mounted on a vibration-isolated optical table.

A variable temperature stage (250 - 400K) was used for these experiments. The resistive heater (25.7 $\Omega$ resistance) was made of constantan alloy wire. The film was glued using silver paste onto the heater top copper plate and its temperature was controlled by a Lakeshore 331 temperature controller using a Platinum temperature sensor. The DC electrical resistance of the sample was measured using a HP34401A multimeter by the 4-wire technique simultaneously with the MOR signals. The heater with the sample was mounted onto a horizontal X-Y-stage driven by two computer-controlled motors, thus allowing the scans to be recorded. Thermal expansions of around 1 $\mu$m / °C in both X and Y directions were measured. The sample was finally placed beneath the objective lens of the MOR system and aligned to the plane perpendicular to the incident beams. At each temperature, the beam focusing onto the film surface was adjusted by varying the height of the objective in order to get maximum lock-in amplifier signal. The sample holder thermal expansion in Z direction was estimated to be around 0.5 $\mu$m / °C.

## 3. RESULTS

Figure 2 shows an area scan recorded at 300 K without any filter in front of the pump laser, thus resulting in a 12.6 mW power coming on the sample. The scan dimensions were 300 µm in X direction (across the bridge) by 790 µm in Y direction (along the bridge). The step size was 5 µm in X direction and 10 µm in Y direction. The MOR signal represented in the (gray) color scale is the voltage directly measured at the output of the lock-in amplifier. We can clearly recognize the geometry of the bridge and the contact pads (schematically represented in the inset of Fig. 2). As expected, no signal resulted from the illumination of the substrate. The central part of the bridge shows quite uniform responses with only 13 % variation, thus potentially demonstrating an overall good homogeneity of the film quality.

Before investigating further the temperature dependence of the MOR signal, we verified the linear dependence of the MOR signal on the pump laser power. The pump laser incident power was varied by inserting filters and figure 3 shows the amplitude of the MOR signal as a function of the power. In the investigated range the MOR signal has a linear response to the pump laser incident power, thus showing that we have not got any saturated phenomena in our measurements.

Secondly, the change of the resistance due to the laser spot illumination (laser spot in the middle of the bridge) was recorded as a function of the temperature and compared to the relative derivative of the resistivity versus temperature, also named resistivity temperature coefficient β and defined as:

$$\beta = 1/\rho \times d\rho/dT \qquad (1)$$

Figure 4-a) shows the electrical resistance change $\Delta r = r_{ON} - r_{OFF}$ when the laser spot position was varied along X defined in the inset. $r_{ON}$ is the resistance of the bridge with the spot on it and $r_{OFF}$ is the resistance with no light. Figure 4-b) shows the temperature dependences of the bridge resistance (left axis), the relative resistance change due the laser spot illumination ($\Delta r/r$) and the coefficient β (right axis). The maximum value of β is 0.018 at 330 K, which is a typical value for the LSMO compound. The optical response was normalized by a constant factor so that the maximums of the curves are the same for easier comparison. One can note the very good superimposition of the curves, revealing the thermal nature of the response, i.e. that the laser does not generate charge carriers and that the temperature change ΔT induced by the laser spot is constant all over the investigated temperature range. We can roughly estimate this local temperature elevation ΔT if we assume that the LSMO film is thermally thin compared to the substrate underneath. In that case the temperature change can be considered dependent on the thermal properties of the substrate material alone, namely $SrTiO_3$. The magnitude of the thermal modulation is given by [17]:

$$\Delta T = \frac{I_0}{2\sqrt{\pi} k_{STO} a} \qquad (2)$$

where $k_{STO}$ is the $SrTiO_3$ thermal conductivity ($k_{STO}$ = 11.35 W.K$^{-1}$.m$^{-1}$), $I_0$ is the incoming optical power ($I_0$ = 3.6 mW) and $a$ is the laser spot radius ($a$ = 7.5 µm). The local temperature variation can then be estimated to be ΔT = 12 K.

Line scans of the reflectance change ΔR across the bridge along X direction are reported in Fig. 5 for different temperatures in the 250 - 400 K range. At each temperature the laser spot was focused on the sample surface so as to get the maximum signal at the output of the lock-in amplifier. The profiles were shifted in the X direction in order to compensate for the thermal expansion. The absolute reflectance $R_0$ of the sample surface at 633 nm was separately measured as a function of the temperature. We obtained a value of $R_0$ = 0.13 at 300K, which is in good agreement with the values reported by Okimoto et al. [18] for single crystals of $La_{2/3}Sr_{1/3}MnO_3$ (0.15 at 2 eV). The average value of each profile line ΔR shown in Fig. 5 was finally divided by the measured reflectivity $R_0$ at each temperature. The resulting values ΔR/$R_0$ are called MOR signal and are reported in Fig. 6, together with the resistivity of the bridge ρ and the relative derivative of the resistivity as a function of the temperature. The MOR signal was normalized by a constant and shifted in temperature by 14 K in order to superimpose the maximums of the curves for easier comparison. The obvious remark is that the two curves have similar shape in the lower temperature part of the graph, and that the required temperature shift is of the order of the previously estimated temperature elevation (about 12 K). This can be simply explained by the fact that the reported temperature is actually the sample holder temperature. The MOR signal is related to the local change of the film reflectance, whereas $1/\rho \times d\rho/dT$ is related to the resistance of the whole bridge and is therefore a global property. For that reason, the observed temperature shift between the two measured quantities is within the expectations.

## 4- ANALYSIS

Several mechanisms can contribute to the change in the optical properties of the manganite. The physics and moreover the electrical conductivity of this material are quite complex and not fully understood yet. The magnitudes of the effects obtained in the present work are significant (a few %) and we discuss below the possible origin of this high MOR response.

The optical properties of manganites have been investigated for various compositions [18,19,20]. A finite-energy peak at low energy is usually observed, revealing a quasi-Drude behavior. Free carriers become dominant in the IR and FIR range and at the wavelength used in this study (633 nm, i.e. 1.96 eV), the reflectivity is likely to be determined by dielectric effects rather than those associated with free carriers. In deriving an expression for the reflectance change however, we will further assume that the temperature variation of the reflectivity can be significantly dependent on carrier concentration because at optical frequencies the dielectric effects are relatively temperature independent. For free carriers, the Drude model expresses the real part of the optical conductivity $\sigma_1$ as:

$$\sigma_1 = \frac{\sigma_0}{1+\omega^2\tau^2} \qquad (3)$$

where $\sigma_0 = \frac{Nq^2\tau}{m^*}$ is equal to the zero frequency electrical conductivity, N is the carrier density, q is the electronic charge, $\tau$ is the scattering rate (mean lifetime of a free carrier between scattering events) and m* is the effective mass of the carriers. The scattering rate therefore writes:

$$\tau = \frac{m^*}{Nq^2\rho} = \frac{1}{\Gamma} \qquad (4)$$

where $\rho$ is the zero frequency electrical resistivity. The plasma oscillation frequency $\omega_P$, which is related to the carrier concentration N, is defined as:

$$\omega_P^2 = \frac{Nq^2}{m^*\varepsilon_0} \qquad (5)$$

The Fresnel equation gives the reflectivity $R(\omega) = \left|\frac{\tilde{n}(\omega)-1}{\tilde{n}(\omega)+1}\right|^2$, where $\omega$ is the scillation frequency of the probe radiation, as a function of the complex refractive index $\tilde{n} = n + jk$. The complex refractive index is related to the dielectric response $\tilde{\varepsilon} = \varepsilon' + j\varepsilon''$ as :

$$\varepsilon' = n^2 - k^2 = 1 + \frac{\omega_P^2 \tau^2}{\omega^2 \tau^2 + 1} \qquad (6)$$

and

$$\varepsilon'' = 2nk = \frac{\omega_P^2 \tau}{\omega(\omega^2 \tau^2 + 1)} \qquad (7)$$

which gives

$$n = \sqrt{\frac{\sqrt{\varepsilon'^2 + \varepsilon''^2} + \varepsilon'}{2}} \qquad (8)$$

and

$$k = \sqrt{\frac{\sqrt{\varepsilon'^2 + \varepsilon''^2} - \varepsilon'}{2}} \qquad (9)$$

In order to estimate the amplitude of the MOR signal one can use the following expression of the fractional change in reflectance from Opsal et al. [2]:

$$\frac{dR}{R} = \text{Re}\left[\frac{4\Delta N}{(\tilde{n}-1)(\tilde{n}+1)} \times \left(\frac{\partial n}{\partial N} + j\frac{\partial k}{\partial N}\right)\right] \qquad (10)$$

where $\Delta N = \frac{\partial N}{\partial T} \times \Delta T$ is the change in carrier concentration N. In the case of YBCO films, equation (10) could be simplified by considering $n \gg k$ and $\varepsilon' \gg \varepsilon''$, and $\omega\tau \gg 1$ [8, 9]. For a rough estimation of the numerical values for LSMO we considered N equal to the nominal carrier (hole) concentration from the formula : 0.3 hole / Mn, which gives $5.6 \times 10^{-27}$ m$^{-3}$ and the effective mass m* equal to $3m_e$ [13]. Table I gives the main calculated parameters for LSMO and YBCO for comparison at 300 K [8, 9]. A first remark should be made about the mean free path value calculated for LSMO at 300 K. It can be estimated using the Mott criterion [21]:

$$l=\frac{\hbar(3\pi^2 N)^{1/3}}{\rho N q^2}=v_F\times\tau \qquad (11)$$

where $v_F$ is the Fermi velocity. The resistivity of our sample at 300 K is 21 µΩ.m. The mean free path is found to be smaller than the interatomic distance (0.3889 nm). This is the signature of charge localization and non coherent metallic phase [13]. Figure 7 gives the evolution of the optical indices, the dielectric response and $\omega\tau$ as function of the temperature. In the considered temperature range, one can reasonably assume that n >> k and ε' >> ε" for writing a simplified variant of the expression (10) and thus giving an estimation of the MOR signal.

Considering n >> k and ε' >> ε" valid, one can write :

$$\frac{dR}{R}\approx\frac{4\Delta N}{(n-1)(n+1)}\times\frac{\partial n}{\partial N} \qquad (12)$$

and determine $\frac{\partial n}{\partial N}$ from equations (6) to (9) as :

$$\frac{\partial n}{\partial N}=\frac{1}{2n}\times\frac{\omega_P^2\tau^2}{N}\times\frac{\omega^2\tau^2-1}{(\omega^2\tau^2+1)^2} \qquad (13)$$

thus leading to the expression of the measured MOR signal as :

$$\frac{\Delta R}{R}=\frac{1}{N}\times\frac{\partial N}{\partial T}\times\frac{2}{\sqrt{1+\frac{\omega_P^2\tau^2}{\omega^2\tau^2+1}}}\times\frac{\omega^2\tau^2-1}{\omega^2\tau^2+1}\times\Delta T \qquad (14)$$

We finally can define a parameter named "dielectric response factor" as:

$$\text{Factor}=\frac{2}{\sqrt{1+\frac{\omega_P^2\tau^2}{\omega^2\tau^2+1}}}\times\frac{\omega^2\tau^2-1}{\omega^2\tau^2+1} \qquad (15)$$

Its temperature dependence is shown in Fig. 7. At sufficiently low temperatures one can consider $\omega\tau$ >> 1 and consequently, the dielectric response factor is independent of the temperature (equation 15). Such approximation leads to a direct relation between the measured MOR signal and the relative change in the carrier concentration $\frac{1}{N} \times \frac{\partial N}{\partial T}$ since $\Delta T$ is constant in the whole investigated temperature range. This feature was deduced from the very good superimposition of the β and Δr/r curves in Fig. 4b. If one could increase ωτ by the use of higher frequency laser radiation, one could get this relation valid up to the temperature values above the Curie temperature, thus expressing the relative variation of the carrier density as a function of the temperature. In Fig. 6 the MOR signal measured when the laser spot was at the centre of the bridge (figure 5) is plotted versus the temperature. The normalized MOR signal was simply shifted by 14 K and divided by 1.79 in order to superimpose the maximum of the MOR signal with the temperature resistance coefficient of the film. No other adjustments were made. We note the qualitative agreement between the two curves below 330 K. The deviation above 330 K might be attributed to variation of the dielectric response factor or to variation of the mobility μ around the metal-to-insulator transition since:

$$\frac{1}{\rho} \times \frac{\partial \rho}{\partial T} = -\left( \frac{1}{\mu} \times \frac{\partial \mu}{\partial T} + \frac{1}{N} \times \frac{\partial N}{\partial T} \right) \tag{16}$$

Figure 6 shows that, in the 260-330 K range, the MOR signal is proportional to the resistivity temperature coefficient. This is a very useful information since it opens the way for several important applications. The MOR technique can be a room temperature, non contact and non destructive method to probe the resistivity variations if the thickness is assumed constant within the film. Secondly, the thickness homogeneity of films can be detected if the resistivity is assumed constant all over the film area. In the latter case the dielectric response factor (equation 15) depends on the film resistivity through the evaluation of τ by equation 4 since the resistivity used in equation 4 is taken from the macroscopic resistance. Figure 8 gives an illustration of this possible application of MOR with the area scans measured for three LSMO films of different average thicknesses that presented a thickness inhomogeneity all over the surface. The dependence of MOR signal with average film thickness suggests that the optical penetration depth of the films exceeds the film thicknesses investigated.

The main advantage of LSMO as functional material is its colossal magnetoresistance. In this study we limited our investigations to the demonstration of the possibility to use MOR as contactless resistance measurement tool via its proportionality to the resistivity temperature coefficient. Experiments where magnetic field is applied and the sample's magnetoresistance is determined by measuring the MOR signal are planned. A further development could be the application of a similar scheme to read information from magnetically recorded media. It would be a new type of reading procedure where the magnetisation of the media will be sensed by local change in the magnetoresistance of a LSMO film placed in close proximity, and this local change in turn will be measured by the MOR technique.

## 5- CONCLUSION

This paper reports MOR measurements performed on CMR thin films. The experimental work shows clear relation between the MOR signal measured as function of the temperature and the relative derivative of the electrical resistivity up to the Curie temperature. Analysis in the framework of the Drude model showed that, within certain conditions, the measured change in reflectivity is correlated to change in the carrier concentration. Measurements at longer wavelength would be beneficial in order to reduce the detrimental variation of other factors versus temperature, and consequently further investigate the metal-to-insulator transition. We have speculated and demonstrated that the MOR technique in combination with CMR films posses a huge potential for applications as contactless resistivity probe – from mapping the resistance or thickness homogeneity of the films to new schemes of reading magnetically recorded media.

Table I. Calculated parameters from above described equations for LSMO and YBCO [8, 9] at 300K.

| | l [nm] | $\tau$ [s] | $\omega\tau$ | $v_F$ [m.s$^{-1}$] | $\omega_P$ [rad.s$^{-1}$] |
|---|---|---|---|---|---|
| LSMO (x=0.3) | 0.19 | 9.0×10$^{-16}$ | 2.7 | 2.1×10$^5$ | 2.4×10$^{15}$ |
| YBCO [8, 9] | 10 | 5.0×10$^{-14}$ | 150 | 2.0×10$^5$ | |

**FIGURE CAPTIONS**

FIG. 1. A schematic diagram of the modulated optical reflectance system.

FIG. 2. Area scan of the 100 μm wide 400 μm long LSMO bridge (pump laser power on the sample surface = 12.6 mW) at room temperature. Inset gives a schematic view of the scanned area.

FIG. 3. MOR signal amplitude (laser spot in the middle of the bridge) as a function of the pump laser incident power. Upper inset shows the measured MOR signals in mV for the three laser powers when the laser spot position was varied along X defined in the lower inset.

FIG. 4. (a) Electrical resistance change $\Delta r = r_{ON} - r_{OFF}$ when the laser spot position was varied along X defined in the inset. $r_{ON}$ stands for the measured resistance value when the spot is scanned along the bridge and $r_{OFF}$ is the value in dark. (b) Resistance versus temperature dependence of the bridge (left axis), relative resistance variation due to the laser spot illumination and resistivity temperature coefficient β versus temperature dependences (right axis).

Figure 5: Reflectance variation ΔR across the bridge (the spot was scanned in X direction) for different temperatures. Curves were normalized in X direction to compensate for the thermal expansion of the sample holder.

FIG. 6. Temperature dependences of the MOR signal ($\Delta R/R_0$), the coefficient β and the resistivity ρ of the bridge. Inset shows the normalized and shifted in X MOR and β curves.

FIG. 7. (a) Evolution of ωτ and of the dielectric response factor (defined in equation 15) as function of the temperature. (b) Evolution of the optical indices and of the dielectric response as function of the temperature.

FIG. 8. Surface scans of the MOR signal measured for three LSMO films of different thicknesses showing a clear non homogeneity in thickness (measured independently with a profilometer).

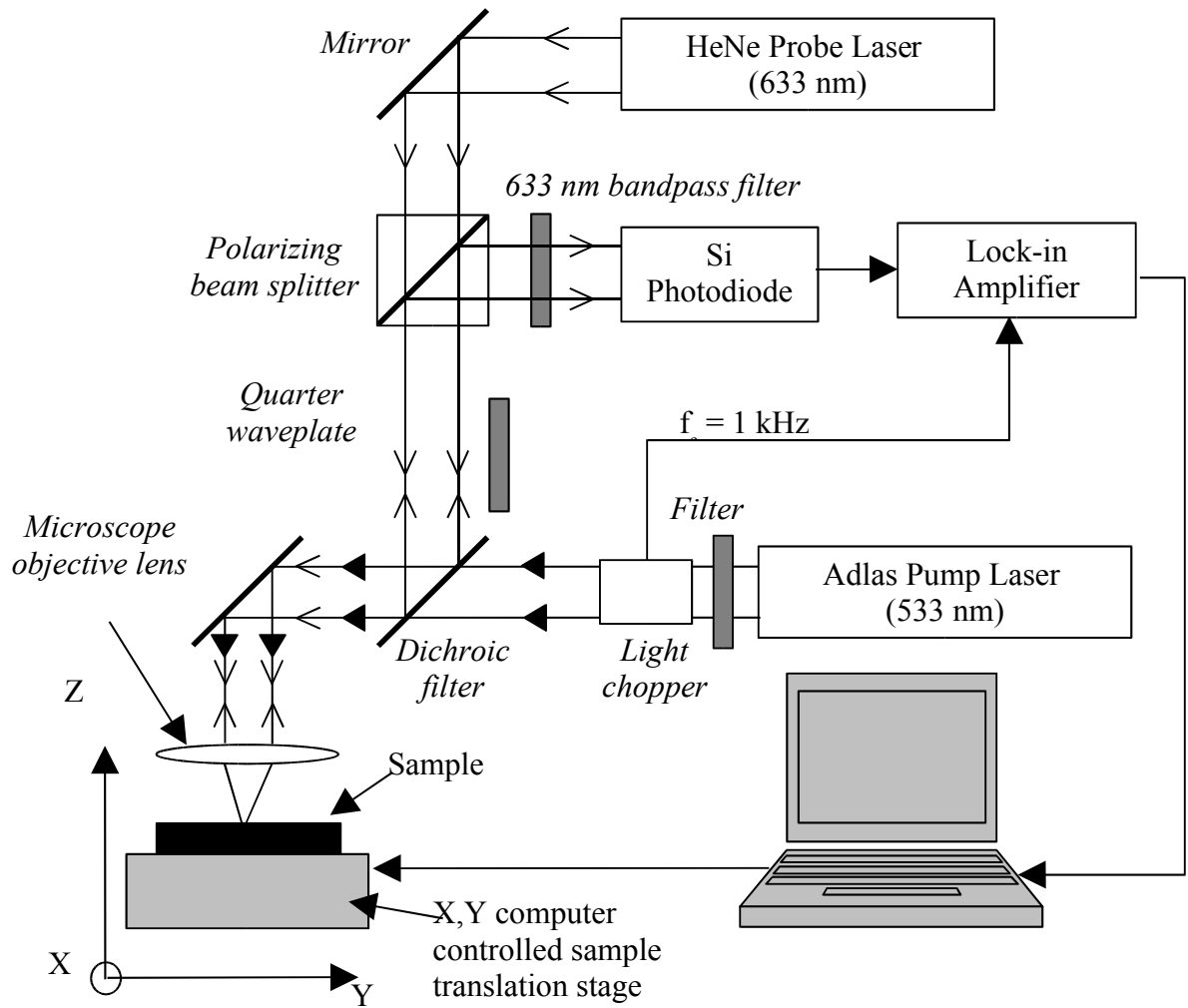

FIG. 1.
MECHIN et al

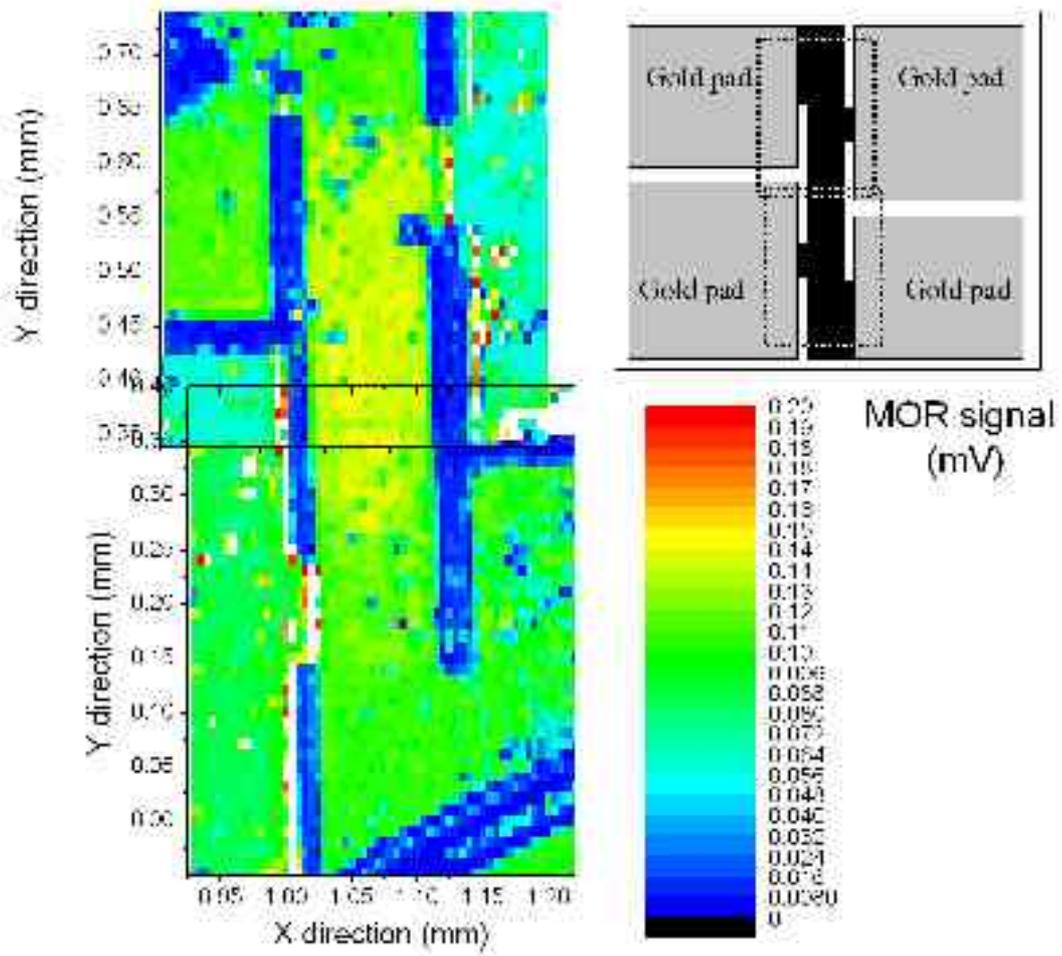

FIG. 2.
MECHIN et al

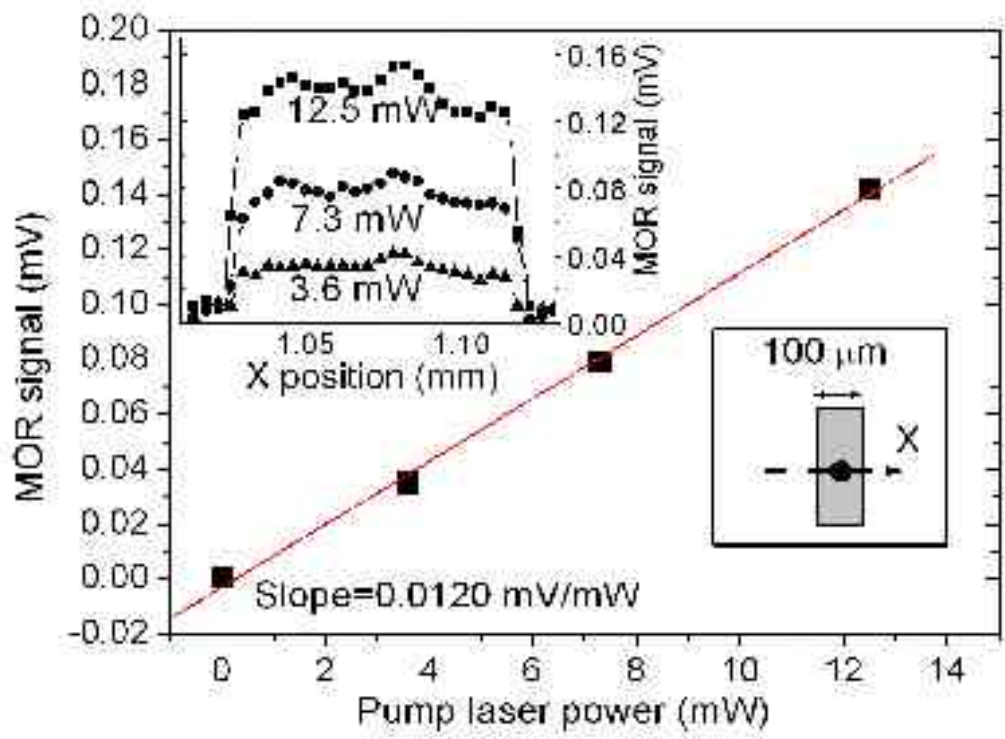

FIG.3.

MECHIN et al

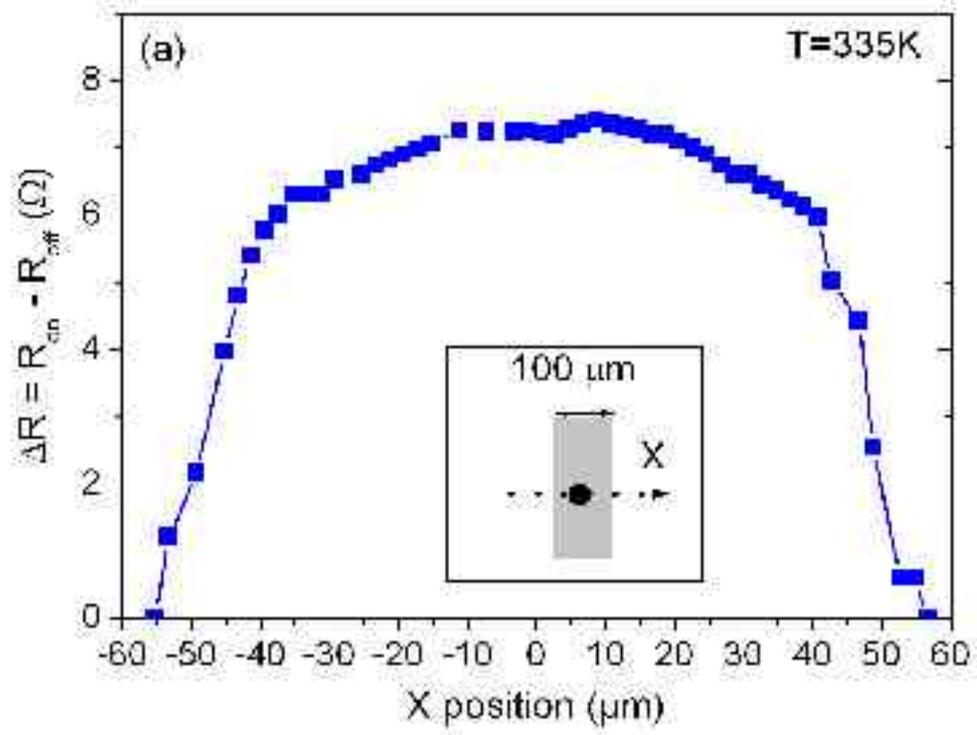

FIG 4 (a)

MECHIN et al

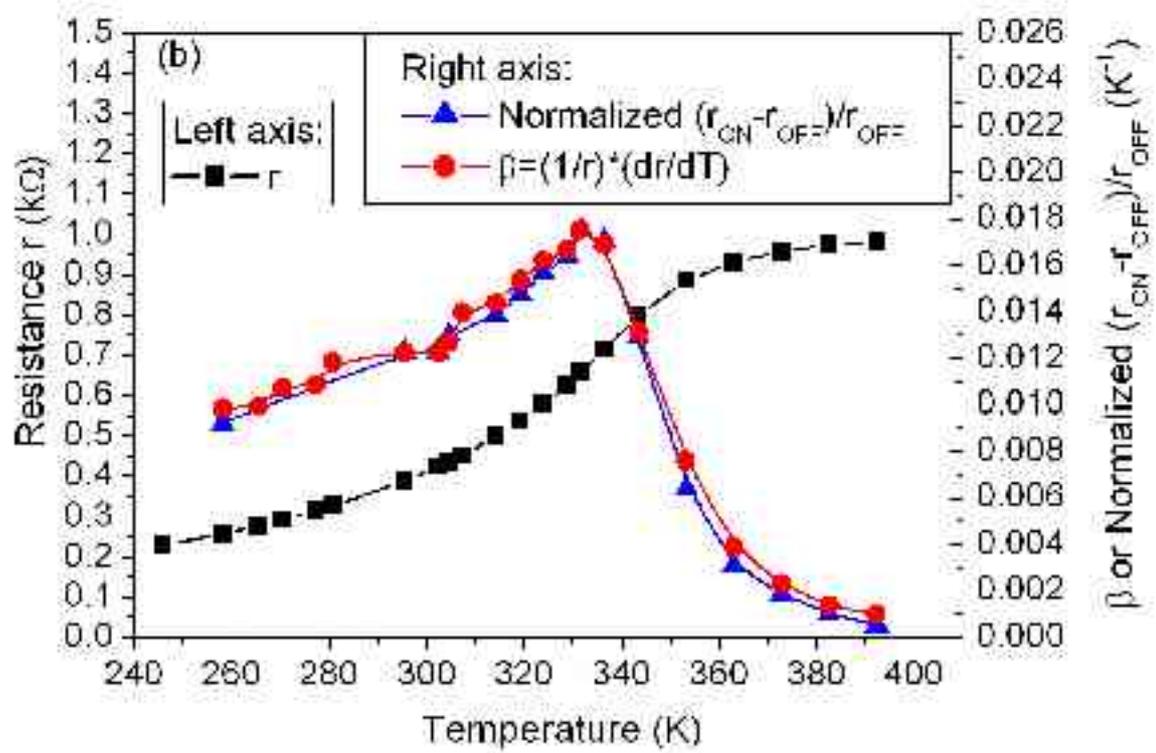

FIG. 4 (b)

MECHIN et al

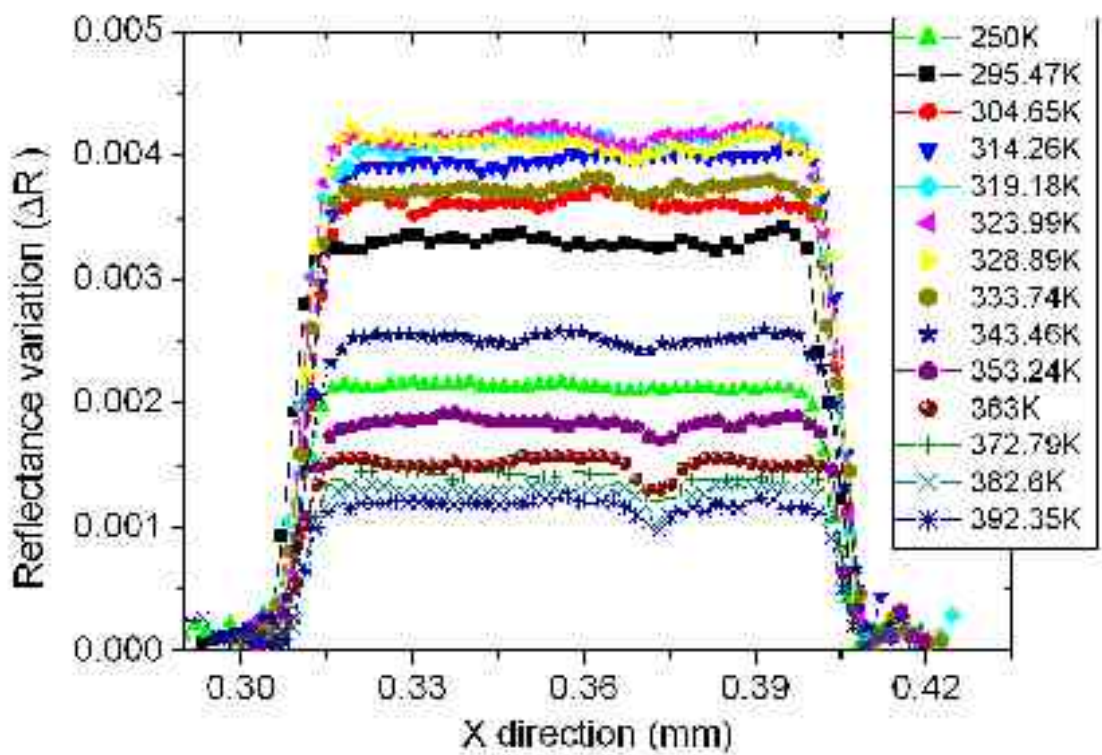

FIG. 5.
MECHIN et al

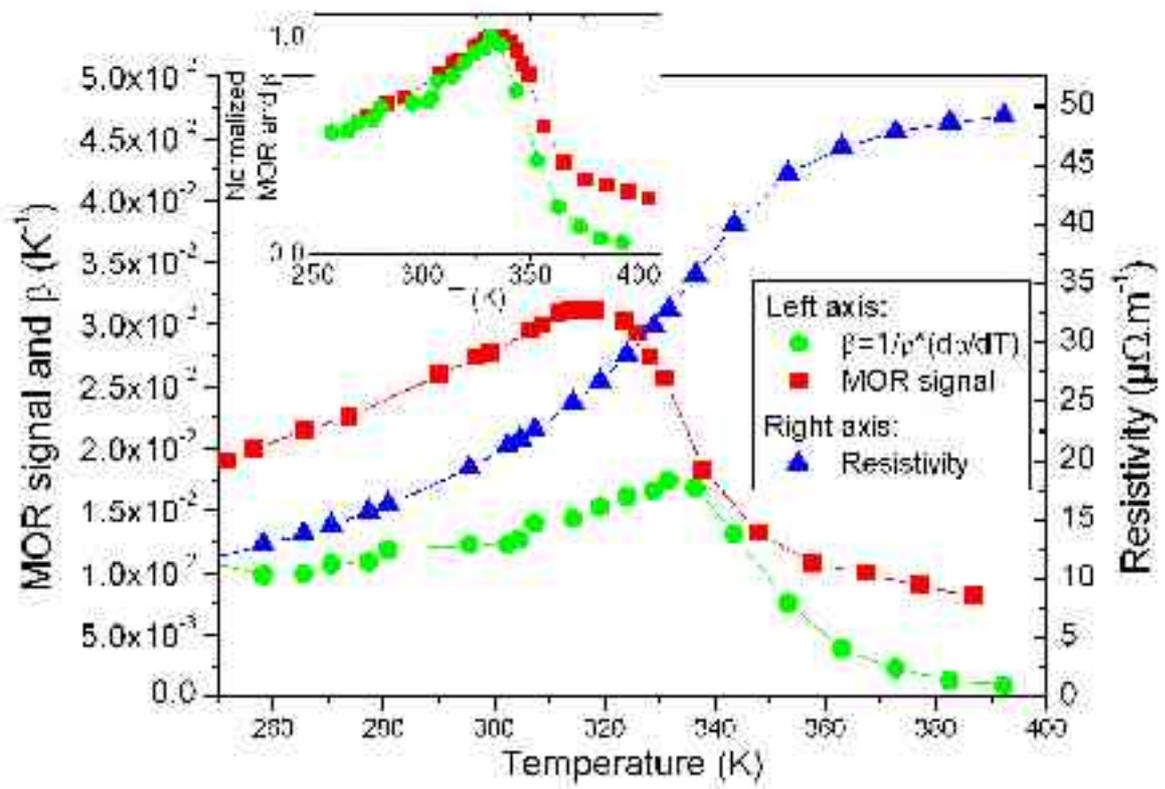

FIG. 6.

MECHIN et al

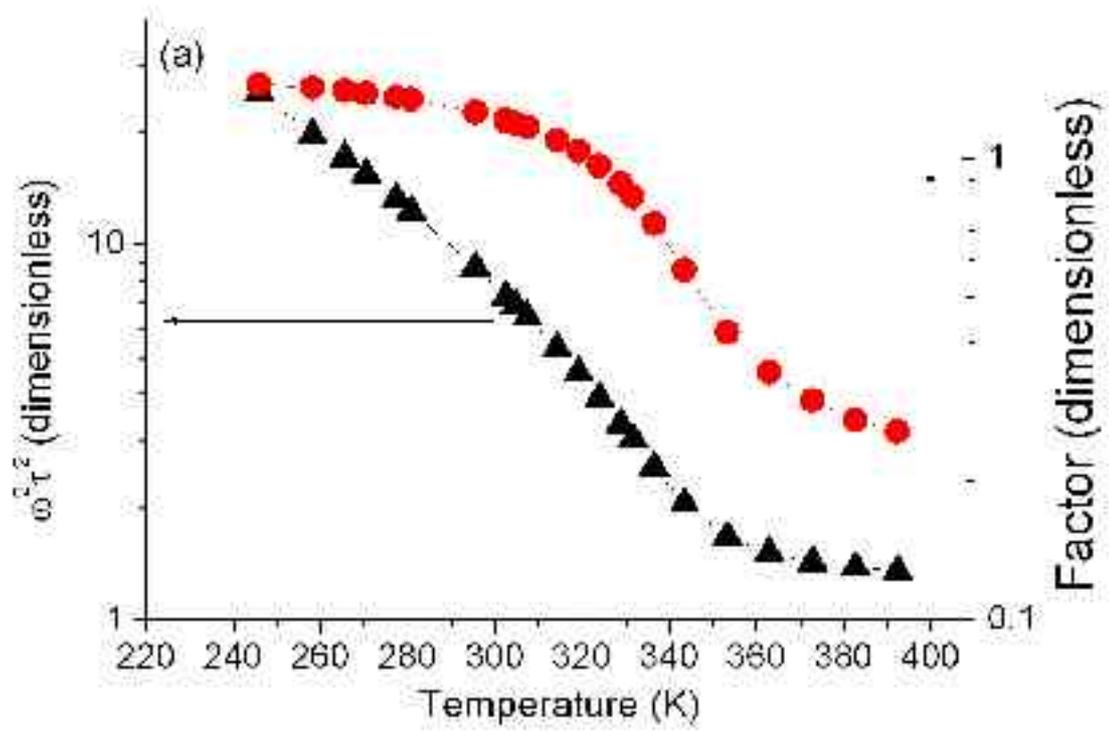

FIG. 7 (a)

MECHIN et al

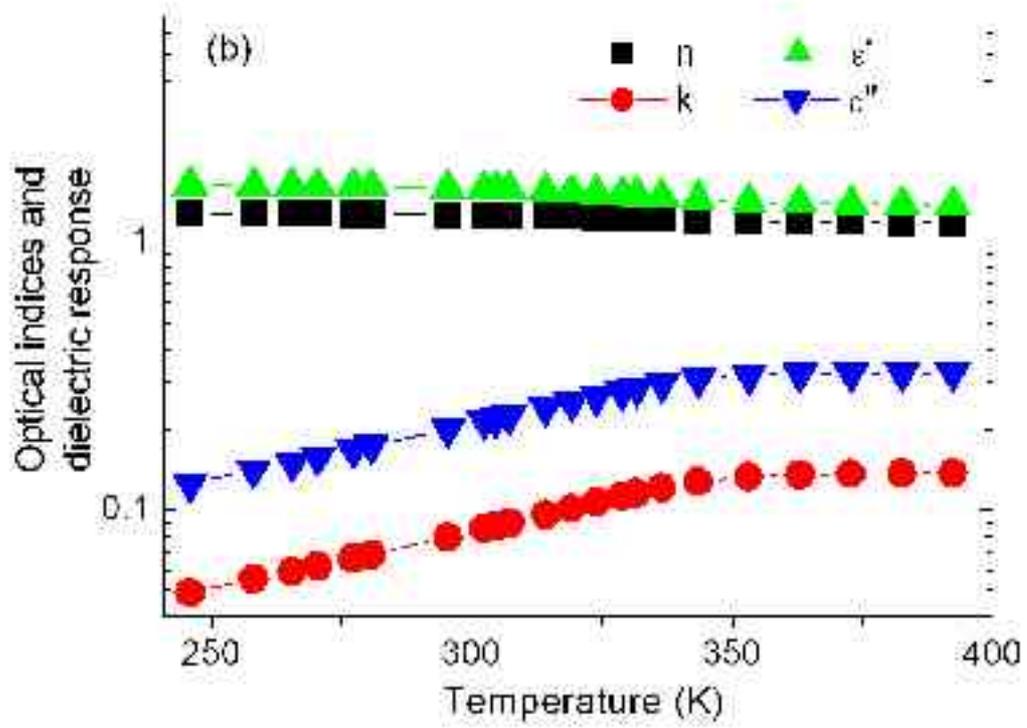

FIG. 7 (b)

MECHIN et al

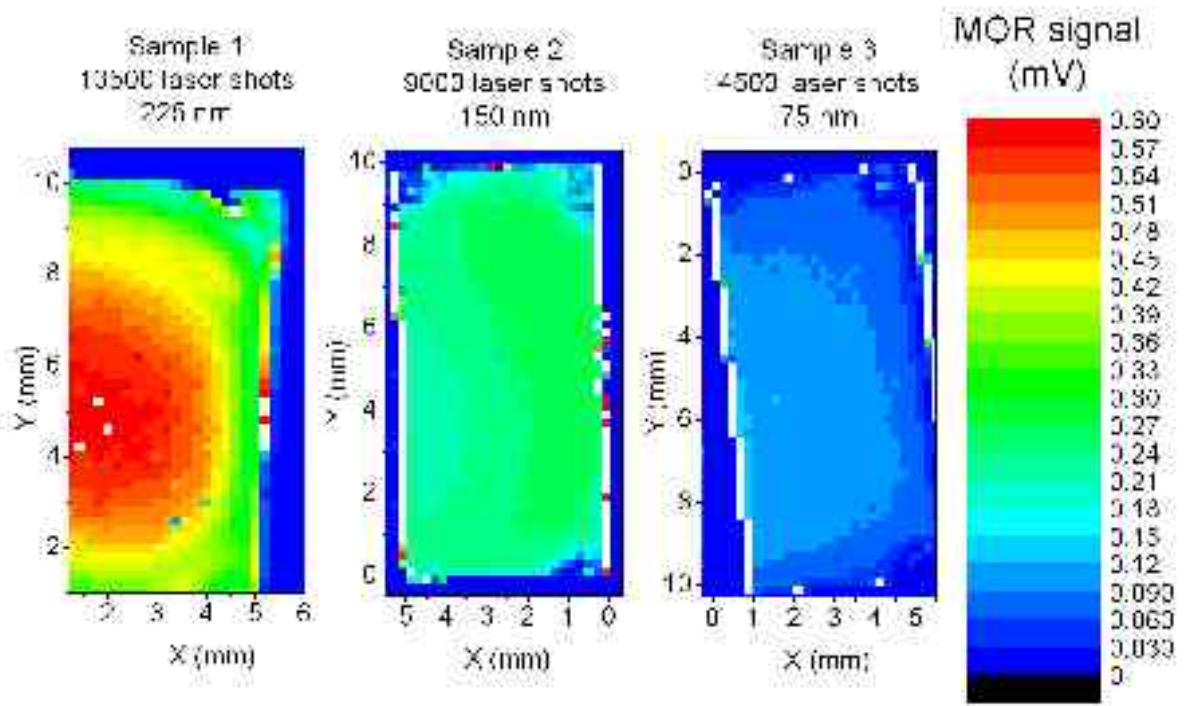

FIG.8.

MECHIN et al